\documentclass[sigconf]{acmart}
\usepackage{booktabs} 

\usepackage{graphicx}        
\usepackage{url}
\usepackage{amsmath,amssymb}
\usepackage{mathtools}
\usepackage{multirow}
\usepackage{subfigure}
\usepackage[ruled, vlined, linesnumbered]{algorithm2e}
\usepackage{float}
\usepackage{color}
\usepackage{bm}
\usepackage{paralist}

\newcommand{\M}[1]{$\mathcal{M}$}

\setcopyright{acmcopyright}

\begin{document}
\title{Data-Driven Modeling of Group Entitativity in Virtual Environments}

\copyrightyear{2018} 
\acmYear{2018} 
\setcopyright{acmcopyright}
\acmConference[VRST '18]{VRST 2018: 24th ACM Symposium on Virtual Reality Software and Technology}{November 28-December 1, 2018}{Tokyo, Japan}
\acmBooktitle{VRST 2018: 24th ACM Symposium on Virtual Reality Software and Technology (VRST '18), November 28-December 1, 2018, Tokyo, Japan}
\acmPrice{15.00}
\acmDOI{10.1145/3281505.3281524}
\acmISBN{978-1-4503-6086-9/18/11}

\author{Aniket Bera}
\affiliation{%
  \institution{University of North Carolina}
   \city{Chapel Hill}
  \state{NC, USA}
  \postcode{27514}
}
\email{ab@cs.unc.edu}

\author{Tanmay Randhavane}
\affiliation{%
  \institution{University of North Carolina}
   \city{Chapel Hill}
  \state{NC, USA}
  \postcode{27514}
}
\email{tanmay@cs.unc.edu}

\author{Emily Kubin}
\affiliation{%
  \institution{Tilburg University}
   \city{North Brabant}
  \state{Netherlands}
  \postcode{27514}
}
\email{e.r.kubin@tilburguniversity.edu}

\author{Husam Shaik}
\affiliation{%
  \institution{University of North Carolina}
   \city{Chapel Hill}
  \state{NC, USA}
  \postcode{27514}
}
\email{hshaik@live.unc.edu}

\author{Kurt Gray}
\affiliation{%
  \institution{University of North Carolina}
   \city{Chapel Hill}
  \state{NC, USA}
  \postcode{27514}
}
\email{kurtgray@unc.edu}

\author{Dinesh Manocha}
\affiliation{%
  \institution{University of Maryland}
   \city{College Park}
  \state{MD, USA}
  \postcode{27514}
}
\email{dm@cs.umd.edu}

\renewcommand{\shortauthors}{Bera et al.}

\begin{abstract}
We present a data-driven algorithm to model and predict the socio-emotional impact of groups on observers. Psychological research finds that highly \textit{entitative} i.e. cohesive and uniform groups induce threat and unease in observers. Our algorithm models realistic trajectory-level behaviors to classify and map the motion-based entitativity of crowds. This mapping is based on a statistical scheme that dynamically learns pedestrian behavior and computes the resultant entitativity induced emotion through group motion characteristics. We also present a novel interactive multi-agent simulation algorithm to model entitative groups and conduct a VR user study to validate the socio-emotional predictive power of our algorithm. We further show that model-generated high-entitativity groups do induce more negative emotions than low-entitative groups. 
\end{abstract}

%
%


\begin{teaserfigure}
  \includegraphics[width=\textwidth]{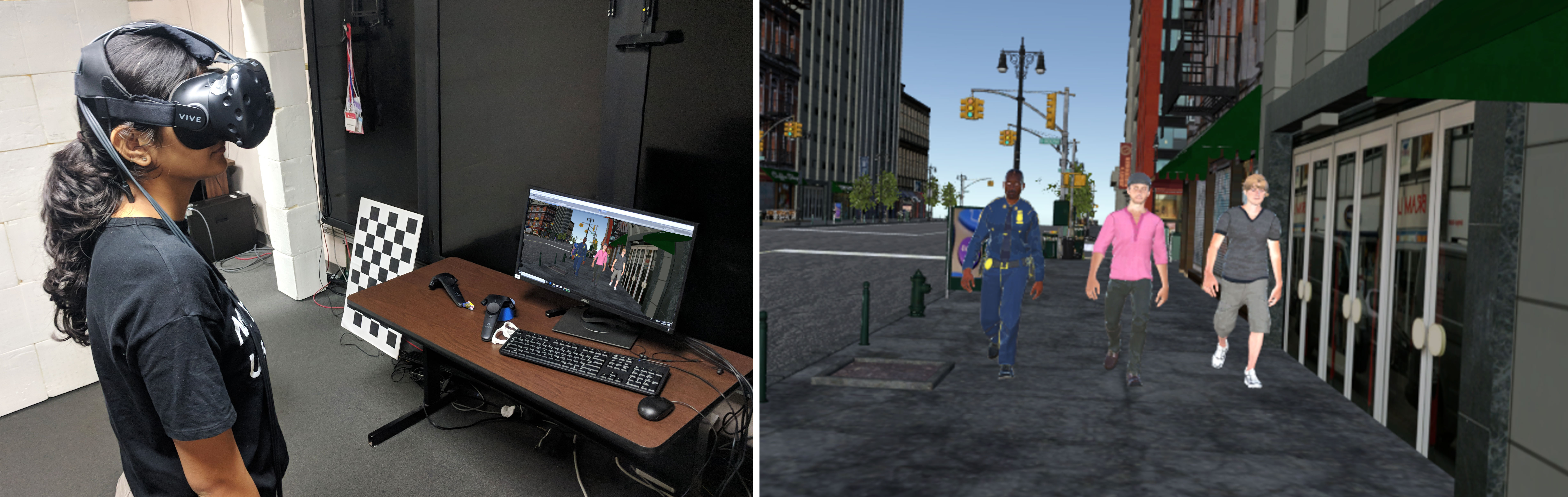}
  \caption{\textbf{Interactive Crowd Simulation:}  Our adaptive data-driven group emotion algorithm is based on a statistical scheme that dynamically learns pedestrian behavior. Our method can generate realistic trajectory-level pedestrian behaviors with varying perceptual entitativity features like  \textit{friendliness, creepiness, comfort}, and \textit{unnervingness}. We can simulate a large number of agent groups at interactive rates.}
  \label{fig:teaser}
\end{teaserfigure}

\maketitle

 \begin{figure*}
  \centering
  \includegraphics[width =1.0 \linewidth]{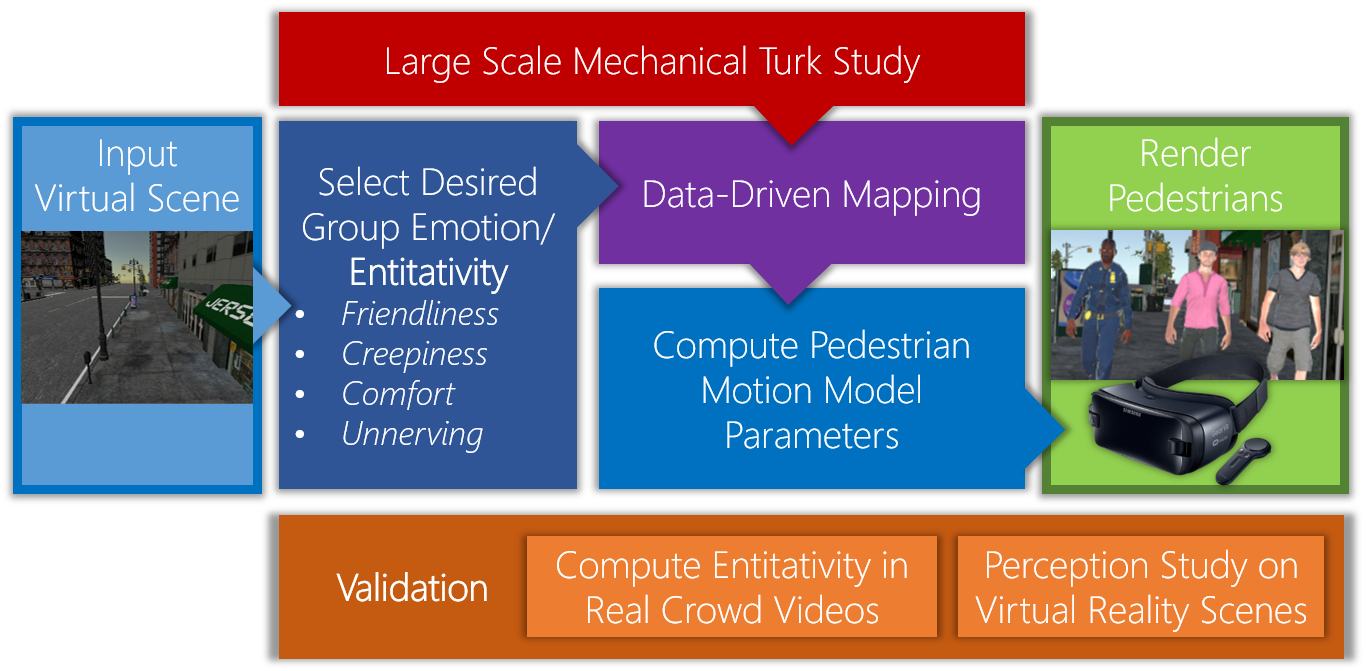}
  \caption{Overview: We highlight the various components of our group emotion-induced entitativity algorithm. We start with a large-scale virtual crowd dataset (generated using a crowd simulation model) and perform a user-perception study to reveal the entitativity level of each group in every video. We present a novel approach to compute the data-driven mapping between entitativity perceptions/features (four inter-related and previously-validated measures of negative socio-affective judgments: \textit{friendliness, creepiness, comfort,} and \textit{unnervingness}) and the motion model parameters which were used to generate the virtual datasets. We then use this data-driven mapping for two use cases: to compute the entitativity in real videos and to simulate human-like characters with varying entitativity. }
  \label{fig:overview}
 \end{figure*}

\section{Introduction}
Understanding and modeling group behavior is an important problem in many domains including virtual reality, robotics, pedestrian dynamics, psychology, and behavior learning. Some of the driving applications include investigation of pathological processes in mental disorders~\cite{diemer2015impact}, virtual reality therapy for crowd phobias~\cite{taffou2014inducing}, training of law enforcement officials or military personnel~\cite{ulicny2001crowd}, understanding crowd flow analysis in urban layouts, etc. In these applications, one of the goals is to generate realistic group movements or emerging behaviors in the background, while the user is immersed in the scene and performing certain tasks. The realism of group movement and the ability to interact with the virtual crowds enhances the presence in the virtual environment and steering strategies~\cite{Hu2013PatternBased}. 

In a group of walking individuals, a critical issue is understanding how the individual motion trajectories combine into ``group-level'' features. Understanding such features is essential because collective and macroscopic motion of groups has the potential to induce a variety of emotional reactions in the user. In a multi-agent environment, some agents may appear friendly whereas some other may appear threatening. These feelings can arise from a variety of sources but frequently stem from how \textit{entitative} a group seems to be. 

\begin{figure}[h]
    \centering
    \includegraphics[width=1.0\linewidth]{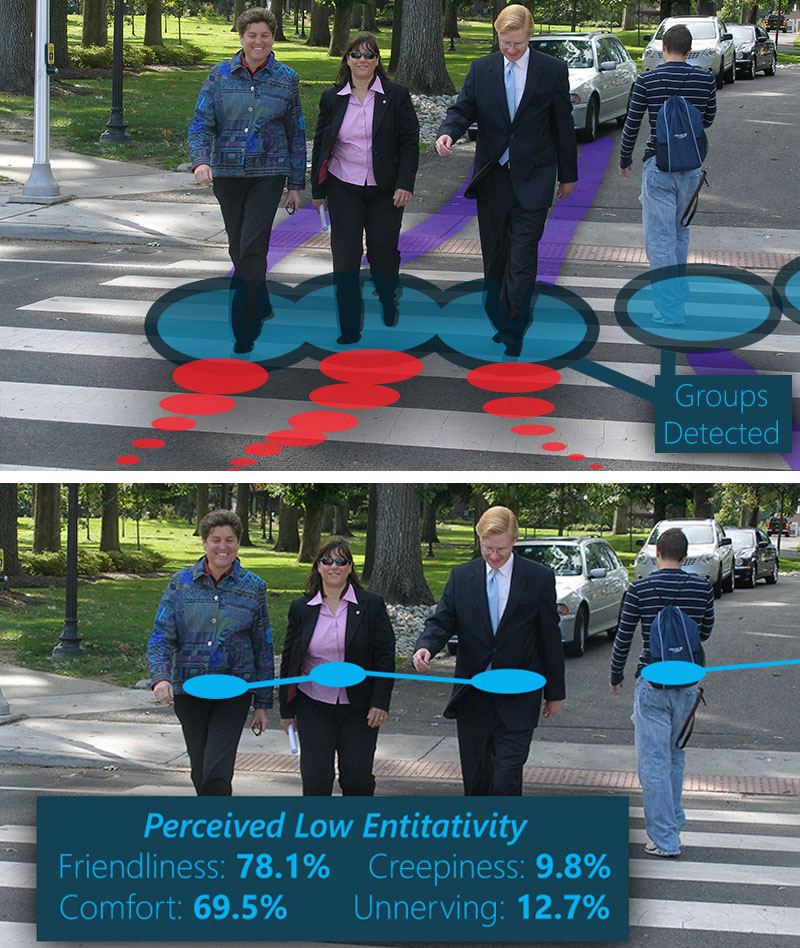}
    \caption{\textbf{Entitativity Classification:} Our novel algorithm can automatically classify the emotion-induction potential of pedestrian group motion based upon \textit{entitativity}. \textbf{Top:} We extract individual pedestrian trajectories from video input, combine these trajectories to cluster individuals into groups. \textbf{Bottom:} Our model then calculates the group's entitativity and its emotion induction potential (operationalized as friendliness, creepiness, unnerving, and comfort).}
   \label{fig:cover}
   \vspace*{-20pt}
\end{figure}

\textit{Entitativity, Emotions, and Reactions}. In this paper, we explore the importance of \textit{entitativity}, which is a measure of how ``group-like'' (i.e., how cohesive, uniform, and similar) a collection of agents seems to be (Figure~\ref{fig:cover}). Entitativity consists of perceptions of similarity, cohesiveness, and uniformity. For example, a military platoon with matching uniforms and haircuts is highly entitative, whereas people waiting at a government office are much less entitative. 

In addition to the similarity of appearance, one important aspect of entitativity is the similarity of movement. When people move together in a cohesive group (with the same trajectory and with proximity) people judge them to be highly ``group-like''. In this paper, we restrict ourselves to only the trajectory-level movement features of the pedestrian groups.

Research in social psychology reveals the importance of group-level characteristics and their impact on others. Since humans are a social species consisting of complex social groups, people are highly sensitive to the characteristics and dynamics of groups ~\cite{pettigrew2013groups}. 
In particular, humans are very attuned to the entitativity of groups because large collections of like-minded people can pose a coordinated threat to others (e.g., gangs, armies; ~\cite{abelson1998perceptions}). Given the potential threat of coordinated groups, when people observe highly entitative groups, having similar appearance and/or motion trajectories, negative emotions are generated. More specifically highly entitative groups have been shown to generate unease and negative socio-emotional appraisals including perceptions/experiences of unfriendliness, creepiness, unnervingness, and discomfort ~\cite{bera2018classifying, bera2018socially}. Therefore, for VR applications including social VR, in order to predict emotions induced by groups of virtual agents, it is necessary to automatically assess how entitative a group appears to be and quickly tie these entitativity assessments to the resultant negative socio-emotional experiences in users. 


\textbf{Main Results:} In this paper, we present a novel data-driven algorithm for modeling and automatically classifying the \textit{entitativity} of a group of pedestrians (Figure~\ref{fig:overview}). Our formulation is based on using the results of an elaborate web-based user study on a large virtual crowd dataset to establish a mapping between trajectory characteristics and their entitativity induced emotions (i.e., socio-emotional appraisals). Consistent with predictions (and past social psychological work)  people report being made more unnerved and uncomfortable by those collections of agents classified by the algorithm as highly entitative. We map the various features of entitativity (operationalized as friendliness, creepiness, unnerving, and comfort) with the motion model parameters that were used to generate the virtual crowd dataset. 

We present two use-cases for our method of being able to \\
a) classify the entitativity of groups in real videos and, \\
b) generate characters in a VR environment to simulate varying degrees of entitativity. 

For the first case, we extract the trajectory of each pedestrian in a video using Bayesian learning at an interactive rate. We cluster the pedestrians in a group and learn various group trajectory-level characteristics. We combine these characteristics to yield an overall entitativity measure based on our mapping, which we tie to negative socio-emotional appraisals. 
For the second case, we generate various virtual reality scenes with virtual agents with varying entitativity. These scenes induce variations of friendliness, creepiness, and comfort in the users.

Finally, we validate the algorithm by performing additional user studies for both the use-cases. The studies reveal that the entitativity algorithm accurately predicts negative socio-emotional appraisals of users.  

Overall, our approach has the following benefits: \\
\noindent {\bf 1. Social Prediction:} Our approach accurately predicts and models important socio-emotional reactions towards other social agents.

\noindent {\bf 2. Robust computation:} Our approach is robust and can account for noise in the pedestrian trajectories.

\noindent {\bf 3. Generalizability:} Our approach is agnostic to the underlying crowd simulation model.

The rest of the paper is organized as follows. In Section~\ref{sec:related}, we review the related work in the field of social psychology and behavior modeling. In Section~\ref{sec:notation}, we give background on quantifying entitativity and introduce our notation. We also present our interactive algorithm, which computes the perceived group entitativity. In Section~\ref{sec:ddem}, we describe our user study on the perception of multiple virtual characters with varying degrees of entitativity. In Section~\ref{sec:real}, we evaluate our algorithm on real videos.In Section~\ref{sec:vr}, we evaluate our simulation algorithm in a VR scene using an HMD.

\section{Related Work}
\label{sec:related}
In this section, we give a brief overview of prior work on psychological perspectives on group dynamics, behavior modeling of pedestrians, and work related to background of crowd analysis.

\subsection{Psychological Perspectives on Group Dynamics}
Human beings are inherently social creatures ~\cite{dunbar1998social}, which evolved to rapidly process and form judgments about collections of people~\cite{saxe2006uniquely}. Many argue the complex social structure and group dynamics of humanity are the key to our ability to be a successful species~\cite{bar2008complexity}. Not only do groups help humans survive by providing benefits that come from collective coordination but they can also represent a definite threat, as groups can make even mild-mannered individuals perpetrate harm upon others~\cite{tajfel1982social}.

Because of the social importance of groups, people have evolved to be extremely adept at detecting information about groups in visual scenes ~\cite{gallagher2009understanding} and rapidly process the social elements that make up groups, such as cohesion and motion.  For example, both classic \cite{heider1944experimental} and modern \cite{gao2009psychophysics} studies find that people automatically see social features in the movement of basic shapes, and people can infer relatively rich social information in groups of people even from a large distance \cite{lee2007group}. Our algorithm is inspired by these classic studies on the rich social context of trajectory motion.  

\subsection{Behavior Modeling of Pedestrians}
There is considerable literature in psychology, robotics, and autonomous driving on modeling the behavior of pedestrians. Many rule-based methods have been proposed to model complex behaviors based on  motor, perceptual, behavioral, and cognitive components~\cite{Terzopoulos}. There is extensive literature on modeling emergent behaviors, starting from Reynolds work~\cite{boids}.  Yeh et al.~\cite{Composite} describe velocity-based methods to model different behaviors including aggression, social priority, authority, protection, and guidance. Other techniques have been proposed to model heterogeneous crowd behaviors based on personality traits ~\cite{GuyPersonality,UPennOCEAN, bera2017aggressive}. Different techniques have been proposed to model collision avoidance behaviors~\cite{lynch2018effect,olivier2018walking} and effect of appearance on perception of virtual characters~\cite{mousas2018effects,zibrek2018effect}.

Our approach is compatible to most of these works and can also predict the entitativity induced emotions of groups of characters created by these methods. 

\subsection{Pedestrian and Crowd Analysis from Videos}
There is extensive work in computer vision and AI literature that analyzes the behaviors and movement patterns of pedestrians in crowd videos~\cite{LiCrowdedSceneAnalysis2015}. The main objectives of these works include human behavior understanding and crowd activity recognition for detecting abnormal behaviors or for surveillance applications~\cite{Hu2004}. Many of these methods use a large number of training videos for offline learning~\cite{EMDPrototype}. Other methods utilize motion models to learn crowd behaviors~\cite{LTACrowd}.

\subsection{Group Dynamics and Socio-Emotional Reaction}

Decades of psychological research reveals that people interact more negatively with groups than with individuals~\cite{quattrone1980perception,yzerbyt1998group}, acting with more hostility towards a group of people rather than a single individual~\cite{cooley2017using}, due to a greater sense of fear~\cite{insko1990individual}. At the heart of anti-social actions are negative socio-emotional reactions, which can be directed at any social agent, whether human, robot~\cite{fraune2015rabble}, or virtual agent~\cite{pertaub2002experiment}. These negative socio-emotional reactions involve appraisals of unease~\cite{bigo2002security}, threat~\cite{fraune2015rabble}, fear~\cite{ommundsen2013exploring} and overall aversive experiences~\cite{strait2015too}.


\section{Notation and Overview}
\label{sec:notation}
In this section, we first define entitativity formally. Then, we introduce the notation and present an overview of the approach.

\subsection{Entitativity}
Entitativity is the perception how much a set of individuals is seen as a single entity (i.e., a group). Perceptions of group entitativity are increased by  perceived psychological similarity, such as when people belong to the same racial groups ~\cite{roets2011role} or ideologies~\cite{harris2011entitativity}, and pursuing the same goal ~\cite{denson2006roles}. Perceptions of group entitativity are also increased by perceived physical similarity, defined as the following  three elements:

\textbf{1. Appearance uniformity:} Highly entitative groups have members that look the same. 

\textbf{2. Common movement: }Highly entitative groups have members that move similarly. 

\textbf{3. Proximity:} Highly entitative groups have members that are very close to each other. 

In this paper, we present an automated algorithm that examines the idea of common movement.  When people move together as a single unit, it leads to perceptions of entitativity which  then induces unease which can in turn lead to aggression and harm. While it is non-trivial to extract or capture the collective pedestrian motion from a video, our approach is based on formulating an entitativity metric from individual trajectories of pedestrians computed using Bayesian learning.

\subsection{Notation and Terminology}
Here, we introduce the notation used in the rest of the paper. We refer to an agent in the crowd as a \textit{pedestrian}. The trajectory and behavior characteristics of each pedestrian are called its {\em state}. These behavior characteristics control how the pedestrian moves on the 2D ground plane.
We refer the pedestrian's state by the symbol $\mathbf{x}\in\mathbb{R}^6$: $\mathbf{x}=[\mathbf{p} \; \mathbf{v}^c \; \mathbf{v}^{pref}]^\mathbf{T}$,
where $\mathbf{p}$ is the pedestrian's position, $\mathbf{v}^c$  is its current velocity, and $\mathbf{v}^{pref}$ is the {\em preferred velocity} on a 2D plane. The preferred velocity is the optimal velocity that a pedestrian would take to achieve its intermediate goal in the absence of other pedestrians or obstacles in the scene. In practice, where other pedestrians and obstacles are present, $\mathbf{v}^{pref}$ tends to be different from $\mathbf{v}^c$ for a given pedestrian. 
The states of all the other pedestrians and the current positions of the obstacles in the scene are collectively represented by the symbol $\mathbf S$. We call this the current state of the environment. We refer to the union of the set of each pedestrian's state as the state of the crowd, which consists of individual pedestrians. We represent this as $\mathbf{X}=\bigcup _i\mathbf{x_i}$, where subscript $i$ denotes the $i^{th}$ pedestrian. In real-world crowds, pedestrians often walk as a part of a group, and we represent a group of pedestrians by $\mathbf{G}=\bigcup _j\mathbf{x_j}$ where subscript $j$ denotes the $j^{th}$ pedestrian in the group. 

Our state formulation does not include any full body or gesture information. Moreover, we do not explicitly model or capture pairwise interactions between pedestrians. However, the difference between $\mathbf{v}^{pref}$ and $\mathbf{v}^c$ provides partial information about the local interactions between a pedestrian and the rest of the environment.

$\mathbf{P}\in\mathbb{R}^6$ denotes the set of parameters for the motion model. The  motion model corresponds to the local navigation rule or scheme that each pedestrian uses to avoid collisions with other pedestrians or obstacles and has a group strategy. Our formulation is based on the RVO velocity-based motion model~\cite{RVO}. In this model, the motion of each pedestrian is governed by these five individual pedestrian characteristics: \textit{Neighbor~Dist, Maximum~Neighbors, Planning~Horizon, (Radius) Personal Space, and Preferred~Speed} and one group characteristic: \textit{Group~Cohesion}. We combine RVO with a group navigation scheme in Section ~\ref{sec:mapping}. In our approach, we mainly analyze four parameters ($\mathbf{GP}\in\mathbb{R}^4$): \textit{Neighbor~Dist, (Radius) Personal Space, Group~Cohesion, and Preferred~Speed}.

{\bf \large Assessing Negative Socio-Emotional Reactions:} 
Prior research in social psychology reveals that entitative groups induce negative  socio-emotional reactions which can license aggression. We therefore use an index of four items to assesses negative  socio-emotional reactions and provide a criterion for our entitativity-induced negative emotions algorithm:
\begin{eqnarray}
\bf{E} = 
 \begin{pmatrix}
Friendliness\\
Creepiness\\
Comfort\\
Unnerving\\
  \end{pmatrix}
\end{eqnarray}

\textit{\textbf{Item and Index Justification}}: Each of these items has been used before in social psychological research in intergroup research to assess negative socio-emotional reactions. Moreover, the high Cronbach's $\alpha$ (a test of statistical reliability) in pilot studies ($\alpha = 0.794$) justifies their combination into a single negative socio-emotional vector.

  \begin{figure*}
  \centering
  \includegraphics[width =1.0 \linewidth]{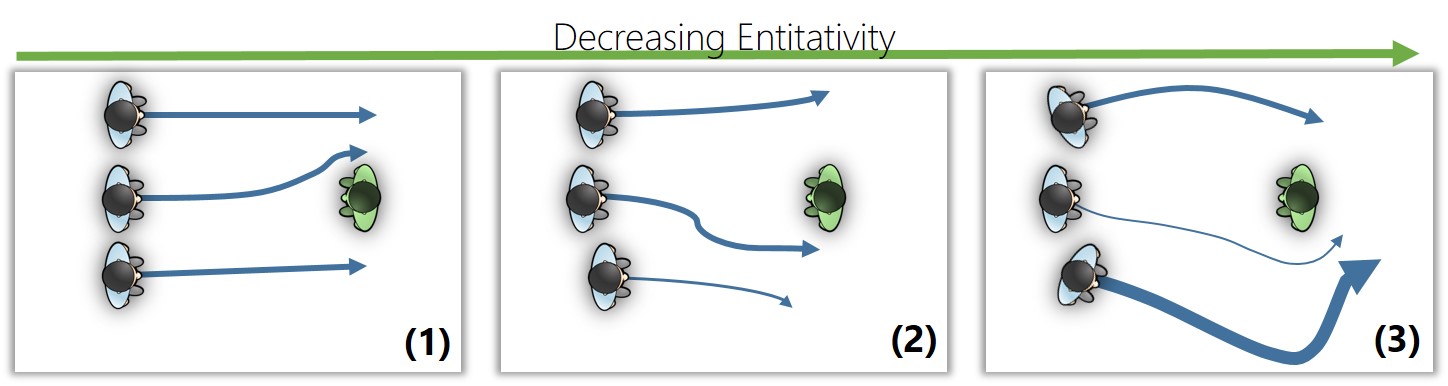}
  \caption{Varying Entitativity - Our user study consisted of three cases for each parameter: (1) high entitativity when there is little to no variation in pedestrian speed and all pedestrian trajectories are somewhat similar/parallel, (2) medium entitativity when there is moderate variation in pedestrian speed and the trajectory directions deviate from each other, (3) low entitativity when there is large variation in pedestrian speed and all pedestrians travel in vastly different directions. The width of the blue arrow denotes the speed (the wider the arrow, the higher the speed) and the direction represents the path taken.}
  \label{fig:responses3}
 \end{figure*}
 
\subsection{Overview}
In this section, we present an overview (Figure \ref{fig:overview}) of our interactive algorithm, which computes the group entitativity model and then simulates virtual agents in real-time.

For each pedestrian in the crowd, the function $G:\mathbb{R}\times\mathbb{R}^6\times\mathbb{S}\rightarrow \mathbb{R}^2$ maps time $t$, the current  state of the pedestrian $\mathbf{x}\in \mathbf{X}$, and the current state of the environment  $\mathbf S \in \mathbb{S}$ to a preferred velocity $\mathbf v^{pref}$. 
Function $I:\mathbb{R}^6\times\mathbb{S}\rightarrow\mathbb{R}^2$ represents the group RVO motion model that is used to compute the current velocity  $\mathbf v^c$  for collision-free interactions with other pedestrians and obstacles. The function $P:\mathbb{R}^2\rightarrow\mathbb{R}^2$ computes the position given $\mathbf v^c$ and $E:\mathbb{R}\rightarrow\mathbb{R}^2$ computes the initial position for time $t_0$ which is the time at which a particular pedestrian enters the environment. 

Each pedestrian uses a local navigation scheme to avoid collisions with other pedestrians and obstacles. In addition to following their individual rules or navigation schemes, pedestrians also navigate as part of a group. We represent these individual as well as group navigation rules by a group motion model. Our formulation is based on a velocity-based motion model which also takes into account proxemic group behaviors~\cite{he2016proxemic}. In this model, the motion of pedestrians is governed by four characteristics/parameters ($\mathbf{GP}$). 

\noindent \textbf{Entitativity Feature Computation}: Once we compute the pedestrian cluster, we make use of a data-driven mapping to compute the collective perceived entitativity-induced emotions $\mathbf{E}$ of the group from the group motion model parameters $\mathbf{GP}$. The derivation and details of this mapping are given in Section~\ref{sec:ddem}.

\section{Data-Driven Entitativity Model}
\label{sec:ddem}
To evaluate the impact of the various parameters of the group motion model on the perception of entitativity of a group of pedestrians, we performed a user study using simulated trajectories. We provide the details of this user study in this section.

\begin{figure}[h]
    \centering
    \includegraphics[width=1.0\linewidth]{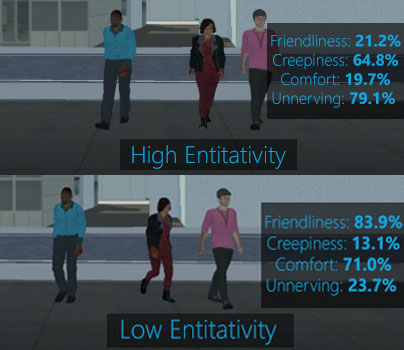}
     \caption{\textbf{Varying Levels of Entitativity:} Parameters of the group motion model affected the entitativity of multiple simulated agents. Agents having the same speed and similar trajectories were perceived to be highly entitative \textit{(top)} whereas agents walking at different speeds and varying trajectories were perceived as less entitative \textit{(bottom)}.}
    \label{fig:varlevel}
\end{figure}

\subsection{Study Goals and Design}
\label{ssec:study}
This study aimed to understand how the perception of multiple pedestrians is affected by the parameters of the group motion model. We use the results of this user study to compute a data-driven statistical mapping between the group motion model parameters and the perception of groups in terms of friendliness, creepiness, and social comfort. We recruited $212$ participants (105 male, 107 female, $\bar{x}_{age}$ = $36$, $s_{age}$ = $11.72$) electronically and also from Amazon MTurk. Here, we provide details of the design of our experiment.









\subsection{Procedure}
\label{ssec:proc}
We performed a web-based study in which the participants were asked to watch pairs of simulated videos of pedestrians and compare the entitativity features (Figure~\ref{fig:varlevel}). Each video contained $3$ simulated agents with various settings of the group motion model parameters. We consider variations in four group motion models parameters ($\mathbf{GP}$): \textit{Neighbor~Dist, Radius, Pref~Speed}, and \textit{Group~Cohesion}. We present the default values for simulation parameters used in our experiments in Table~\ref{tab:defaultval}. In each pair, one of the videos corresponds to the default values of the parameters (called the \textit{Reference} video). We generated the other video (called the \textit{Question} video) by varying one parameter to either the minimum or the maximum value. Thus a total of $8$ pairs of videos were generated corresponding to the minimum and the maximum value for each motion model parameter. Out of the $8$ videos, each participant watched a random subset of $4$ pairs of videos. The participants watched the video pairs side by side in randomized order. They could watch the videos multiple times if they wished and compared the entitativity features of the pedestrian groups in the two videos. We collected the demographic information about participants' gender and age at the end of the study.

\subsection{Questions} 
\label{ssec:ques}
For each trial, the participant compared the two videos (\textit{Reference} and \textit{Question}) on a 6-point scale from Strongly Disagree (1) - Strongly Agree (6). The following items were adapted  ~\cite{gray2012feeling} to assess creepiness and social comfort experienced: 

As compared to the \textit{Reference} video, in the \textit{Question} video ...
\begin{compactitem}
\item Did the characters seem more friendly?
\item Did the characters seem more creepy?
\item Did you feel more comfortable with the characters?
\item Did you feel more unnerved by the characters?
\end{compactitem}
These questions were motivated by previous studies~\cite{quattrone1980perception}. We define an entitativity feature corresponding to each question. Thus, we represent the entitativity features of a group as a 4-D vector: \textit{Friendliness, Creepiness, Comfort, Unnerving (Ability to Unnerve)}.

\subsection{Analysis} \label{sec:mapping}
We average the participant responses to each video pair to obtain $8$ entitativity feature data points ($\textbf{E}_i, i = 1, 2, ..., 8\}$). The range of entitativity features in these data points is presented in Table~\ref{tab:dataRange}. We also present the standard deviation of the features.  

Table~\ref{tab:correl} provides the correlation coefficients between the features for all the participant responses. The high correlation between the features indicates that the features measure different aspects of socio-emotional reactions to entitativity. As expected, \textit{creepiness} and \textit{unnerving} are inversely correlated with \textit{friendliness} and \textit{comfort}. Principal Component Analysis of the four socio-emotional features also reveals that a single principal component is enough to explain over 98\% of the variance in the participants' responses. We use this component to combine the four entitativity features into an entitativity label $e_{i} \in \mathbb{R}$: 
\begin{multline}
e_{i} = -0.31*\textbf{E}_i^{Friendliness} + 0.66*\textbf{E}_i^{Creepiness} \\ - 0.46*\textbf{E}_i^{Comfort} + 0.51*\textbf{E}_i^{Unnerving} \label{eq:pcaEntit}
\end{multline}

\begin{table}[h]
\begin{center}
 
     \begin{tabular}{|l|l|l|l|}
     \hline
     \textbf{Parameters ($\mathbf{GP}$)}                 & \textbf{min} & \textbf{max} & \textbf{default} \\ \hline
     Neighbor Distance ($m$)     & 3   & 5  & 4   \\
     Radius (Personal Space) ($m$)                & 0.8 & 1.7 & 1.0  \\
     Preferred speed ($m/s$)    & 1.2 & 1.8   & 1.5    \\
     Group Cohesion & 0.1 & 1.0   & 0.5    \\ \hline
    \end{tabular}
    \caption{Default values for simulation parameters used in our experiments}
    \label{tab:defaultval}
\end{center}
\end{table}

\begin{table}[h]
\centering
\begin{tabular}{|c|c|c|c|}
\hline
 & Min & Max & STD \\ \hline
Friendliness & 2.664 & 3.636 & 0.392 \\ \hline
Creepiness & 2.654 & 4.452 & 0.797 \\ \hline
Comfort & 2.617 & 3.810 & 0.557 \\ \hline
Unnerving & 2.882 & 4.343 & 0.631 \\ \hline
\end{tabular}
\caption{\textbf{Range of the Entitativity Features}: For low and high values of motion model parameters (keeping the other parameters at default value) we obtain the above range of entitativity features.}
\label{tab:dataRange}
\end{table} 

\begin{table}[h]
\centering
\resizebox{\linewidth}{!}{
\begin{tabular}{|c|c|c|c|c|}
\hline
 & Friendliness & Creepiness & Comfort & Unnerving \\ \hline
Friendliness & 1 & -0.963 &	0.973 &	-0.944 \\ \hline
Creepiness & -0.963 & 1 & -0.990 &	0.977 \\ \hline
Comfort & 0.973 & -0.990 & 1 & -0.969 \\ \hline
Unnerving & -0.944 & 0.977 & -0.969 & 1 \\ \hline
\end{tabular}}
\caption{\textbf{Correlation Between Questions}: We provide the correlation coefficients between the questions. The high correlation between the questions indicates that the questions measure different aspects of a single perception feature, entitativity.}
\label{tab:correl}
\end{table}

We normalize the responses and obtain entitativity values ($e_i, i = 1, 2, ..., 8\}$) for each variation of the motion model parameters ($\textbf{GP}_i, i = 1, 2, ..., 8\}$), we can fit a linear model to the entitativity and the model parameters. We refer to this model as the \textit{Data-Driven Entitativity Model}. For each video pair $i$ in the gait dataset, we have a vector of parameter values $\textbf{GP}_i$ and an entitativity value $e_i$. Given these parameters and features, we compute the entitativity mapping of the form:

\begin{multline}
	e = a_0 + a_1 * Neighbor~Dist + a_2 * Radius
	\\ + a_3 * Pref.~Speed + a_4 * Group~Cohesion
\end{multline}
We obtain the coefficient vector $A = \{a_0, a_1, a_2, a_3, a_4\}$ using linear regression with entitativity values as the responses and the parameter values as the predictors using the normal distribution ($R^2 = 0.942, F(1, 3)=29.6, p < 0.01$):
\begin{equation}
 \mathbf{A} = [0.60\quad -0.42\quad -0.58\quad 0.75\quad 0.73]
\end{equation}

We can make many inferences from the values of $\bf{A}$. The negative values of $a_0$ and $a_1$ indicate that as the values of neighbor distance and radius increases, the entitativity of the group decreases. That is, groups with larger interpersonal distances appear less entitative. This validates the psychological findings in the previous literature. Entitativity increases with walking speed and group cohesion. This indicates that faster-walking groups of agents appear discomforting and less friendly.

We can use our data-driven entitativity model to predict perceived entitativity of any group for any new input video. Given the motion parameter values $\mathbf{GP}$ for the group, the perceived entitativity $e$ can be obtained as: $e = \bf{A}*\bf{GP}$.

In addition to computing entitativity from the motion parameters, we can also predict individual features of entitativity as well. We perform multiple linear regression and obtain the following equation:
\begin{eqnarray}
\resizebox{.45 \textwidth}{!}{
$\bf{E} = 
\begin{bmatrix}
0.32 & 0.20 & 0.33 & -0.23 & -0.42 \\
0.56 & -0.45 & -0.51 & 0.73 & 0.69 \\
0.34 & 0.28 & 0.36 & -0.49 & -0.59 \\
0.57 & -0.29 & -0.52 & 0.68 & 0.47
\end{bmatrix}
\begin{bmatrix}
1\\ 
Neighbor\, Dist\\ 
Radius\\ 
Pref\, Speed\\ 
Group\, Cohesion
\end{bmatrix}
$
}\label{eq:individualMapping}
\end{eqnarray}


Given the motion model parameters $\mathbf{GP}$, we can compute any of the features of entitativity ($\mathbf{E}$ = [\textit{Friendliness, Creepiness, Comfort, Unnerving}]) using Equation~\ref{eq:individualMapping}. $R^2$ and F-statistic values in Table~\ref{tab:modelFitting} indicate that our linear model fits the data well.

\begin{table}[]
\centering
\begin{tabular}{|c|c|c|c|}
\hline
             & $R^2$ & F    & p     \\ \hline
Friendliness & 0.980                & 85.0 & 0.002 \\ \hline
Creepiness   & 0.916                & 20.1 & 0.017 \\ \hline
Comfortable  & 0.933                & 25.2 & 0.012 \\ \hline
Unnerving    & 0.952                & 35.8 & 0.007 \\ \hline
\end{tabular}
\caption{$R^2$ and F-statistic values which indicate that our model fits the data well.}
\label{tab:modelFitting}
\vspace*{-25pt}
\end{table}

\section{Validation on Real Data}
\label{sec:real}
We validated our algorithm on videos of pedestrians walking in the real world. We performed a user study to obtain the values of perceived entitativity for our validation videos. We computed the error between the user reported entitativity values and the values computed using our Data-Driven Entitativity Model. We describe the details of the user study below. We also applied our novel algorithms to the 2D pedestrian trajectories generated and extracted from different publicly available crowd videos and calculated the performance.

\noindent \textbf{Pedestrian Tracking}: Our method takes a crowd video as an input and we extract the initial set of pedestrian trajectories using an online pedestrian tracker. We learn the pedestrian group motion model parameters using statistical methods~\cite{berareach, beraglmp}. We use the Bayesian-inference technique to compensate for any errors and to compute the state of each pedestrian. We use an Ensemble Kalman Filter (EnKF) and Expectation Maximization (EM) to estimate the most likely state $\mathbf{x}$ of each pedestrian. Our approach extends the method presented in~\cite{kim2016interactive}. 


\subsection{Study Goals}
This study aimed to obtain the perceived entitativity values for videos of pedestrians walking in the real world. We recruited $46$ participants (29 male, 17 female, $\bar{x}_{age}$ = $32.52$, $s_{age}$ = $10.34$) electronically and also from Amazon MTurk. Participants watched the videos in the dataset and answered some questions. We presented the videos in randomized order to the participants, and they could watch each video as many times as they wanted. After answering the questions for all the videos, participants provided demographic information. We asked questions which are similar to Section~\ref{ssec:study} and were adapted~\cite{gray2012feeling} to assess creepiness and social comfort experienced. 

\subsection{Dataset}
\begin{figure*}
    \centering
    \includegraphics[width=1.0\linewidth]{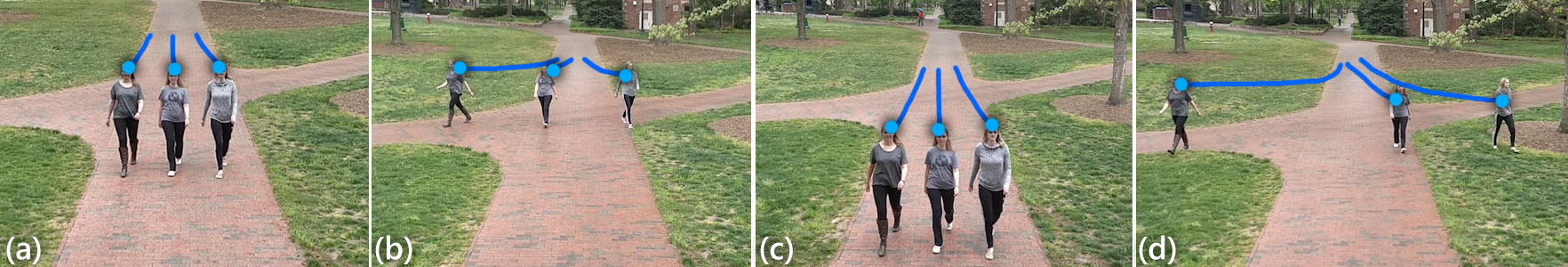}
    \caption{We captured videos of pedestrians walking in the real world. In each of these videos (\textasciitilde $10$ seconds), three pedestrians were walking on a university campus.}
    \label{fig:RealVideos}
\end{figure*}

We captured eight videos of pedestrians walking in the real world. In each of these videos (\textasciitilde $10$ seconds), three pedestrians were walking on a university campus (Figure ~\ref{fig:RealVideos}). We used the videos which contained the same set of pedestrians to account for differences in appearance and body shapes.

\subsection{Results and Analysis}
Using the participant responses, we calculated entitativity values using a method similar to Section~\ref{sec:mapping}. We averaged the participant responses and combined the four entitativity features using the obtained PCA coefficients. We obtained the entitativity values $e_i \in \mathbb{R}$ for each video $i$ using Equation~\ref{eq:pcaEntit}. 

Here, $\textbf{E}_i^{Friendliness}$, $\textbf{E}_i^{Creepiness}$, $\textbf{E}_i^{Comfort}$, $\textbf{E}_i^{Unnerving}$
are the average participant responses to the four questions. We normalized these entitativity values $e_i^{ground}$ which we treat as ground truth for error computation. For each video, we also computed the pedestrian parameters $\textbf{GP}$. Given $\textbf{GP}$, we used $e$ (Section~\ref{ssec:proc}) to compute the predicted entitativity value $e_i^{pred}$ for each video $i$. We define the $error(e_i^{ground}, e_i^{pred})$ between the ground truth $e_i^{ground}$ and predicted entitativity $e_i^{pred}$  as:

\begin{eqnarray}
    error(e_i^{ground}, e_i^{pred}) = \frac{|e_i^{ground} - e_i^{pred}|}{e_{max} - e_{min}}
\end{eqnarray}
where $e_{max}$ and $e_{min}$ are the maximum and minimum attainable entitativity values. For our dataset, we observe an average error of $5.91\%$ which indicates that the perceived entitativity of groups of real pedestrians by participants matches the predicted entitativity by our algorithm.

\section{Validation in a VR Environment}
\label{sec:vr}
We validated our algorithm on videos of pedestrians walking in a VR setting. 

\subsection{Study Goal}
We conducted a within users VR study to evaluate the perception of entitativity of a group of pedestrians in a virtual environment. This study aimed to validate whether our entitativity model correctly predicts the socio-emotional reactions of the users.

\subsection{Experimental Design}
We conducted the user study using an HTC VIVE HMD on a desktop machine with an Nvidia Titan X GPU, Intel Xeon E5-1620 v3 4-core processor, 16 GB of memory, and Windows 10 OS. The study involved $4$ scenarios with each scenario containing $3$ characters. Each scenario had different entitativity parameters related to a group of three virtual pedestrians (Table~\ref{tab:scenarios}).

\begin{table}[h]
\begin{center}
    \begin{tabular}{ | l | p{4.5cm} | p{1.5cm} |}
    \hline
    \textbf{Scenarios} & \textbf{Description} & \textbf{Entitativity Level} \\ \hline
   \textbf{ Scene 1} &  Pedestrians walked with identical trajectories, high synchronization and movement, and high cohesion. & Highest \\ \hline
    \textbf{Scene 2} & Pedestrians had the \textbf{Scene 1} trajectories and movement with a slightly lower cohesion, less synchronized and slightly slower walking pace. & High \\ \hline
    \textbf{Scene 3} &  Pedestrians walked with varying trajectories and movement with low cohesion at a regular walking pace. & Medium\\
    \hline
    \textbf{Scene 4} &  Pedestrians had the \textbf{Scene 3} trajectories and movement with a lower cohesion and a slower walking pace. & Low \\
    \hline
    \end{tabular}
    \caption{Four scenarios with varying levels of entitativity were used to validate our algorithm in a virtual environment.}
    \label{tab:scenarios}
\end{center}
\end{table}

\subsection{Participants}
Participants were recruited from staff and students of a university. Total 30 participants (17 males, 12 females) were recruited.

\subsection{Procedure}
The participants were asked to watch scenarios with a group of virtual pedestrians with varying entitativity in a virtual environment and evaluate their perception (friendly/creepy) of the group. 

Each participant was presented with the 4 scenarios on the HTC VIVE HMD in a randomized order. The participants were free to look and walk around the environment, but there was no interaction between the virtual pedestrians in the scenes and the participant. The participants rated their perception of the pedestrian group using questions similar to Section IV on a 4-point scale from Strongly Disagree(1) to Strongly Agree(4). 

\subsection{Results}
We provide the mean values of participant responses to the four questions for the four scenes in Table~\ref{tab:meanVR}. Mean values of friendliness and comfort decrease with higher entitativity whereas mean values of creepiness and unnervingness increase.

We combined the participant responses to obtain the entitativity labels using Equation~\ref{eq:pcaEntit}. There was a statistically significant difference between the responses for the different levels of entitativity as determined by one-way ANOVA $(F(3,30) = 249.44, p < 0.001)$. We also conducted pairwise t-tests to compare the four entitativity levels. 

\subsection{Discussion}
Our results (Table~\ref{tab:ttest}) indicate that there was a significant difference between all the pairwise comparisons. Our algorithm can create different levels of entitativity that can induce more negative emotions than low-entitative groups.

Since realistic behaviors, interactions, and movements of virtual agents can increase the sense of presence and immersion in VR, understanding user-crowd interaction is important. As shown by our results, varying entitativity can be used to create virtual experiences having characters with different levels of friendliness, creepiness, comfort, and unnervingness. 


\begin{table}[h]
\begin{center}
 \resizebox{\linewidth}{!}{
    \begin{tabular}{|l|l|l|l|l|}
    \hline
    \textbf{Scene} & \textbf{Friendliness} & \textbf{Creepiness} & \textbf{Comfort} & \textbf{Unnerving} \\ \hline
    Scene 1 & 1.30	& 3.47	& 1.47	& 3.60 \\
    Scene 2 & 1.57	& 3.20	& 1.73	& 3.27 \\
    Scene 3 & 2.83	& 2.07	& 3.20	& 2.00 \\
    Scene 4 & 3.63	& 1.47	& 3.67	& 1.23 \\ \hline
    \end{tabular}}
    \caption{Mean participant responses to the VR scenes.}
    \label{tab:meanVR}
    \vspace*{-25pt}
\end{center}
\end{table}

\begin{table}[h]
\begin{center}
 
    \begin{tabular}{|l|l|l|}
    \hline
    \textbf{Scenes} & \textbf{t(29)} & \textbf{p} \\ \hline
    Scene 1 and 2 & 4.38	& <0.001\\
    Scene 1 and 3 & 17.62	& <0.001\\
    Scene 1 and 4 & 21.25	& <0.001\\
    Scene 2 and 3 & 13.57	& <0.001\\ 
    Scene 2 and 4 & 17.59	& <0.001\\
    Scene 3 and 4 & 8.80	& <0.001\\\hline
    \end{tabular}
    \caption{Results of our pairwise t-tests.}
    \label{tab:ttest}
    \vspace*{-25pt}
\end{center}
\end{table}



\section{Conclusions, Limitations, and Future Work}
We present a novel method for automatically classifying  \textit{entitativity} via pedestrian motion trajectories.  
Our algorithm identifies groups of pedestrians and extracts their individual trajectories before synthesizing them together into  group-level characteristics. We precompute a data-driven entitativity metric that predicts negative socio-emotional reactions (supported by a VR user study) with important implications for real-world behavior. To the best of our knowledge, this is the first approach for automatic video-based group entitativity classification and the associated prediction of socio-emotional reactions. Our approach can also be used in VR applications including social VR to create scenarios with groups of virtual agents having varying entitativity levels which can induce different levels of friendliness, comfort, and creepiness in the users.

Our approach has some limitations. Most importantly, people rely upon more than group-level motion characteristics when making judgments of groups of people, also relying upon rich social and identity information (e.g., perceptions of race, class, religion, and gender). Our algorithm only considers socio-emotional reactions to motion trajectories and so may not capture the other diverse inputs of socio-emotional reactions. Additionally, the predictive benefit of our algorithm may be limited in some cases, such as in a very large high-density group in which motion-trajectories are heavily constrained.

A key future direction involves extending the prediction of entitativity judgment to additional cues, primarily appearance. When individuals look more similar based on race, clothing, or posture, they are seen to be more entitative. This trajectory-based algorithm could be expanded to incorporate appearance-based cues, such as overall color. An additional future direction would be to understand entitativity judgments as a combination of both the behavior of others and the personality of the perceiver.

\section{Acknowledgements}
This research is supported in part by ARO grant W911NF16-1-0085, and Intel. We thank Austin Wang for his help in designing 3D scenes for the user study and Tanya Amert for lending her voice for the video.

\bibliographystyle{unsrtnat}
\bibliography{references}

\begin{thebibliography}{48}
\providecommand{\natexlab}[1]{#1}
\providecommand{\url}[1]{\texttt{#1}}
\expandafter\ifx\csname urlstyle\endcsname\relax
  \providecommand{\doi}[1]{doi: #1}\else
  \providecommand{\doi}{doi: \begingroup \urlstyle{rm}\Url}\fi

\bibitem[Diemer et~al.(2015)Diemer, Alpers, Peperkorn, Shiban, and
  M{\"u}hlberger]{diemer2015impact}
Julia Diemer, Georg~W Alpers, Henrik~M Peperkorn, Youssef Shiban, and Andreas
  M{\"u}hlberger.
\newblock The impact of perception and presence on emotional reactions: a
  review of research in virtual reality.
\newblock \emph{Frontiers in psychology}, 6, 2015.

\bibitem[Taffou(2014)]{taffou2014inducing}
Marine Taffou.
\newblock \emph{Inducing feelings of fear with virtual reality: the influence
  of multisensory stimulation on negative emotional experience}.
\newblock PhD thesis, Paris 6, 2014.

\bibitem[Ulicny and Thalmann(2001)]{ulicny2001crowd}
Branislav Ulicny and Daniel Thalmann.
\newblock \emph{Crowd simulation for interactive virtual environments and VR
  training systems}.
\newblock Springer, 2001.

\bibitem[Hu et~al.(2013)Hu, Lees, and Zhou]{Hu2013PatternBased}
Nan Hu, Michael~Harold Lees, and Suiping Zhou.
\newblock A pattern-based modeling framework for simulating human-like
  pedestrian steering behaviors.
\newblock In \emph{Proceedings of the 19th ACM Symposium on Virtual Reality
  Software and Technology}, VRST '13, pages 179--188, New York, NY, USA, 2013.
  ACM.
\newblock ISBN 978-1-4503-2379-6.
\newblock \doi{10.1145/2503713.2503723}.
\newblock URL \url{http://doi.acm.org/10.1145/2503713.2503723}.

\bibitem[Pettigrew and Tropp(2013)]{pettigrew2013groups}
Thomas~F Pettigrew and Linda~R Tropp.
\newblock \emph{When groups meet: The dynamics of intergroup contact}.
\newblock Psychology Press, 2013.

\bibitem[Abelson et~al.(1998)Abelson, Dasgupta, Park, and
  Banaji]{abelson1998perceptions}
Robert~P Abelson, Nilanjana Dasgupta, Jaihyun Park, and Mahzarin~R Banaji.
\newblock Perceptions of the collective other.
\newblock \emph{Personality and Social Psychology Review}, 2\penalty0
  (4):\penalty0 243--250, 1998.

\bibitem[Bera et~al.(2018{\natexlab{a}})Bera, Randhavane, Kubin, Wang, Manocha,
  and Gray]{bera2018classifying}
Aniket Bera, Tanmay Randhavane, Emily Kubin, Austin Wang, Dinesh Manocha, and
  Kurt Gray.
\newblock Classifying group emotions for socially-aware autonomous vehicle
  navigation.
\newblock In \emph{Proceedings of the IEEE Conference on Computer Vision and
  Pattern Recognition Workshops}, pages 1039--1047, 2018{\natexlab{a}}.

\bibitem[Bera et~al.(2018{\natexlab{b}})Bera, Randhavane, Kubin, Wang, Gray,
  and Manocha]{bera2018socially}
Aniket Bera, Tanmay Randhavane, Emily Kubin, Austin Wang, Kurt Gray, and Dinesh
  Manocha.
\newblock The socially invisible robot: Navigation in the social world using
  robot entitativity.
\newblock In \emph{Intelligent Robots and Systems (IROS)}, 2018{\natexlab{b}}.

\bibitem[Dunbar(1998)]{dunbar1998social}
RI~Dunbar.
\newblock The social brain hypothesis.
\newblock \emph{brain}, 9\penalty0 (10):\penalty0 178--190, 1998.

\bibitem[Saxe(2006)]{saxe2006uniquely}
Rebecca Saxe.
\newblock Uniquely human social cognition.
\newblock \emph{Current opinion in neurobiology}, 16\penalty0 (2):\penalty0
  235--239, 2006.

\bibitem[Bar-Yam(2008)]{bar2008complexity}
Yaneer Bar-Yam.
\newblock Complexity rising: From human beings to human civilization, a
  complexity profile, 2008.

\bibitem[Tajfel(1982)]{tajfel1982social}
Henri Tajfel.
\newblock Social psychology of intergroup relations.
\newblock \emph{Annual review of psychology}, 33\penalty0 (1):\penalty0 1--39,
  1982.

\bibitem[Gallagher and Chen(2009)]{gallagher2009understanding}
Andrew~C Gallagher and Tsuhan Chen.
\newblock Understanding images of groups of people.
\newblock In \emph{Computer Vision and Pattern Recognition, 2009. CVPR 2009.
  IEEE Conference on}, pages 256--263. IEEE, 2009.

\bibitem[Heider and Simmel(1944)]{heider1944experimental}
Fritz Heider and Marianne Simmel.
\newblock An experimental study of apparent behavior.
\newblock \emph{The American journal of psychology}, 57\penalty0 (2):\penalty0
  243--259, 1944.

\bibitem[Gao et~al.(2009)Gao, Newman, and Scholl]{gao2009psychophysics}
Tao Gao, George~E Newman, and Brian~J Scholl.
\newblock The psychophysics of chasing: A case study in the perception of
  animacy.
\newblock \emph{Cognitive psychology}, 59\penalty0 (2):\penalty0 154--179,
  2009.

\bibitem[Lee et~al.(2007)Lee, Choi, Hong, and Lee]{lee2007group}
Kang~Hoon Lee, Myung~Geol Choi, Qyoun Hong, and Jehee Lee.
\newblock Group behavior from video: a data-driven approach to crowd
  simulation.
\newblock In \emph{Proceedings of the 2007 ACM SIGGRAPH/Eurographics symposium
  on Computer animation}, pages 109--118. Eurographics Association, 2007.

\bibitem[Shao and Terzopoulos(2005)]{Terzopoulos}
Wei Shao and Demetri Terzopoulos.
\newblock Autonomous pedestrians.
\newblock In \emph{Symposium on Computer animation}, pages 19--28, 2005.
\newblock ISBN 1-59593-198-8.

\bibitem[Reynolds(1999)]{boids}
Craig Reynolds.
\newblock {Steering Behaviors for Autonomous Characters}.
\newblock In \emph{Game Developers Conference 1999}, 1999.

\bibitem[Yeh et~al.(2008)Yeh, Curtis, Patil, van~den Berg, Manocha, and
  Lin]{Composite}
H.~Yeh, S.~Curtis, S.~Patil, J.~van~den Berg, D.~Manocha, and M.~Lin.
\newblock Composite agents.
\newblock In \emph{Symposium on Computer Animation}, pages 39--47, 2008.
\newblock ISBN 978-3-905674-10-1.

\bibitem[Guy et~al.(2011)Guy, Kim, Lin, and Manocha]{GuyPersonality}
Stephen~J. Guy, Sujeong Kim, Ming~C. Lin, and Dinesh Manocha.
\newblock Simulating heterogeneous crowd behaviors using personality trait
  theory.
\newblock In \emph{Symposium on Computer Animation}, pages 43--52. ACM, 2011.
\newblock ISBN 978-1-4503-0923-3.

\bibitem[Durupinar et~al.(2011)Durupinar, Pelechano, Allbeck, G{\"u}~andd{\"u}
  andkbay, and Badler]{UPennOCEAN}
F.~Durupinar, N.~Pelechano, J.M. Allbeck, U.~G{\"u}~andd{\"u} andkbay, and N.I.
  Badler.
\newblock How the ocean personality model affects the perception of crowds.
\newblock \emph{Computer Graphics and Applications, IEEE}, 31\penalty0
  (3):\penalty0 22 --31, may-june 2011.
\newblock ISSN 0272-1716.

\bibitem[Bera et~al.(2017)Bera, Randhavane, and Manocha]{bera2017aggressive}
Aniket Bera, Tanmay Randhavane, and Dinesh Manocha.
\newblock Aggressive, tense, or shy? identifying personality traits from crowd
  videos.
\newblock In \emph{Proceedings of the Twenty-Sixth International Joint
  Conference on Artificial Intelligence, IJCAI-17}, pages 112--118, 2017.

\bibitem[Lynch et~al.(2018)Lynch, Pettr{\'e}, Bruneau, Kulpa, Cr{\'e}tual, and
  Olivier]{lynch2018effect}
Sean~D Lynch, Julien Pettr{\'e}, Julien Bruneau, Richard Kulpa, Armel
  Cr{\'e}tual, and Anne-Helene Olivier.
\newblock Effect of virtual human gaze behaviour during an orthogonal collision
  avoidance walking task.
\newblock In \emph{2018 IEEE Conference on Virtual Reality and 3D User
  Interfaces (VR)}, pages 136--142. IEEE, 2018.

\bibitem[Olivier et~al.(2018)Olivier, Bruneau, Kulpa, and
  Pettr{\'e}]{olivier2018walking}
Anne-H{\'e}l{\`e}ne Olivier, Julien Bruneau, Richard Kulpa, and Julien
  Pettr{\'e}.
\newblock Walking with virtual people: Evaluation of locomotion interfaces in
  dynamic environments.
\newblock \emph{IEEE transactions on visualization and computer graphics},
  24\penalty0 (7):\penalty0 2251--2263, 2018.

\bibitem[Mousas et~al.(2018)Mousas, Anastasiou, and
  Spantidi]{mousas2018effects}
Christos Mousas, Dimitris Anastasiou, and Ourania Spantidi.
\newblock The effects of appearance and motion of virtual characters on
  emotional reactivity.
\newblock \emph{Computers in Human Behavior}, 86:\penalty0 99--108, 2018.

\bibitem[Zibrek et~al.(2018)Zibrek, Kokkinara, and McDonnell]{zibrek2018effect}
Katja Zibrek, Elena Kokkinara, and Rachel McDonnell.
\newblock The effect of realistic appearance of virtual characters in immersive
  environments-does the character's personality play a role?
\newblock \emph{IEEE transactions on visualization and computer graphics},
  24\penalty0 (4):\penalty0 1681--1690, 2018.

\bibitem[Li et~al.(2015)Li, Chang, Wang, Ni, Hong, and
  Yan]{LiCrowdedSceneAnalysis2015}
Teng Li, Huan Chang, Meng Wang, Bingbing Ni, Richang Hong, and Shuicheng Yan.
\newblock Crowded scene analysis: A survey.
\newblock \emph{Circuits and Systems for Video Technology, IEEE Transactions
  on}, 25\penalty0 (3):\penalty0 367--386, March 2015.

\bibitem[Hu et~al.(2004)Hu, Tan, Wang, and Maybank]{Hu2004}
Weiming Hu, Tieniu Tan, Liang Wang, and Steve Maybank.
\newblock A survey on visual surveillance of object motion and behaviors.
\newblock \emph{Systems, Man, and Cybernetics, Part C: Applications and
  Reviews, IEEE Transactions on}, 34\penalty0 (3):\penalty0 334--352, 2004.

\bibitem[Zen and Ricci(2011)]{EMDPrototype}
G.~Zen and E.~Ricci.
\newblock Earth mover's prototypes: A convex learning approach for discovering
  activity patterns in dynamic scenes.
\newblock In \emph{Computer Vision and Pattern Recognition (CVPR), 2011 IEEE
  Conference on}, pages 3225--3232, June 2011.
\newblock \doi{10.1109/CVPR.2011.5995578}.

\bibitem[Pellegrini et~al.(2012)Pellegrini, Gall, Sigal, and Gool]{LTACrowd}
Stefano Pellegrini, Jürgen Gall, Leonid Sigal, and Luc Gool.
\newblock Destination flow for crowd simulation.
\newblock In \emph{Computer Vision – ECCV 2012. Workshops and
  Demonstrations}, volume 7585, pages 162--171. 2012.
\newblock ISBN 978-3-642-33884-7.
\newblock \doi{10.1007/978-3-642-33885-4_17}.
\newblock URL \url{http://dx.doi.org/10.1007/978-3-642-33885-4_17}.

\bibitem[Quattrone and Jones(1980)]{quattrone1980perception}
George~A Quattrone and Edward~E Jones.
\newblock The perception of variability within in-groups and out-groups:
  Implications for the law of small numbers.
\newblock \emph{Journal of Personality and Social Psychology}, 38\penalty0
  (1):\penalty0 141, 1980.

\bibitem[Yzerbyt et~al.(1998)Yzerbyt, Rogier, and Fiske]{yzerbyt1998group}
Vincent~Y Yzerbyt, Anouk Rogier, and Susan~T Fiske.
\newblock Group entitativity and social attribution: On translating situational
  constraints into stereotypes.
\newblock \emph{Personality and Social Psychology Bulletin}, 24\penalty0
  (10):\penalty0 1089--1103, 1998.

\bibitem[Cooley and Payne(2017)]{cooley2017using}
Erin Cooley and B~Keith Payne.
\newblock Using groups to measure intergroup prejudice.
\newblock \emph{Personality and Social Psychology Bulletin}, 43\penalty0
  (1):\penalty0 46--59, 2017.

\bibitem[Insko et~al.(1990)Insko, Schopler, Hoyle, Dardis, and
  Graetz]{insko1990individual}
Chester~A Insko, John Schopler, Rick~H Hoyle, Gregory~J Dardis, and Kenneth~A
  Graetz.
\newblock Individual-group discontinuity as a function of fear and greed.
\newblock \emph{Journal of Personality and Social Psychology}, 58\penalty0
  (1):\penalty0 68, 1990.

\bibitem[Fraune et~al.(2015)Fraune, Sherrin, Sabanovi{\'c}, and
  Smith]{fraune2015rabble}
Marlena~R Fraune, Steven Sherrin, Selma Sabanovi{\'c}, and Eliot~R Smith.
\newblock Rabble of robots effects: Number and type of robots modulates
  attitudes, emotions, and stereotypes.
\newblock In \emph{Proceedings of the Tenth Annual ACM/IEEE International
  Conference on Human-Robot Interaction}, pages 109--116. ACM, 2015.

\bibitem[Pertaub et~al.(2002)Pertaub, Slater, and
  Barker]{pertaub2002experiment}
David-Paul Pertaub, Mel Slater, and Chris Barker.
\newblock An experiment on public speaking anxiety in response to three
  different types of virtual audience.
\newblock \emph{Presence: Teleoperators \& Virtual Environments}, 11\penalty0
  (1):\penalty0 68--78, 2002.

\bibitem[Bigo(2002)]{bigo2002security}
Didier Bigo.
\newblock Security and immigration: Toward a critique of the governmentality of
  unease.
\newblock \emph{Alternatives}, 27\penalty0 (1):\penalty0 63--92, 2002.

\bibitem[Ommundsen et~al.(2013)Ommundsen, Yakushko, Van~der Veer, and
  Ulleberg]{ommundsen2013exploring}
Reidar Ommundsen, Oksana Yakushko, Kees Van~der Veer, and P{\aa}l Ulleberg.
\newblock Exploring the relationships between fear-related xenophobia,
  perceptions of out-group entitativity, and social contact in norway.
\newblock \emph{Psychological reports}, 112\penalty0 (1):\penalty0 109--124,
  2013.

\bibitem[Strait et~al.(2015)Strait, Vujovic, Floerke, Scheutz, and
  Urry]{strait2015too}
Megan Strait, Lara Vujovic, Victoria Floerke, Matthias Scheutz, and Heather
  Urry.
\newblock Too much humanness for human-robot interaction: exposure to highly
  humanlike robots elicits aversive responding in observers.
\newblock In \emph{Proceedings of the 33rd Annual ACM Conference on Human
  Factors in Computing Systems}, pages 3593--3602. ACM, 2015.

\bibitem[Roets and Van~Hiel(2011)]{roets2011role}
Arne Roets and Alain Van~Hiel.
\newblock The role of need for closure in essentialist entitativity beliefs and
  prejudice: An epistemic needs approach to racial categorization.
\newblock \emph{British Journal of Social Psychology}, 50\penalty0
  (1):\penalty0 52--73, 2011.

\bibitem[Harris(2011)]{harris2011entitativity}
Kira~J Harris.
\newblock Entitativity and ideology: a grounded theory of disengagement.
\newblock 2011.

\bibitem[Denson et~al.(2006)Denson, Lickel, Curtis, Stenstrom, and
  Ames]{denson2006roles}
Thomas~F Denson, Brian Lickel, Mathew Curtis, Douglas~M Stenstrom, and Daniel~R
  Ames.
\newblock The roles of entitativity and essentiality in judgments of collective
  responsibility.
\newblock \emph{Group Processes \& Intergroup Relations}, 9\penalty0
  (1):\penalty0 43--61, 2006.

\bibitem[van~den Berg et~al.(2008)van~den Berg, Lin, and Manocha]{RVO}
J.~van~den Berg, Ming Lin, and D.~Manocha.
\newblock Reciprocal velocity obstacles for real-time multi-agent navigation.
\newblock In \emph{Robotics and Automation, 2008. ICRA 2008. IEEE International
  Conference on}, pages 1928 --1935, may 2008.

\bibitem[He et~al.(2016)He, Pan, Wang, and Manocha]{he2016proxemic}
Liang He, Jia Pan, Wenping Wang, and Dinesh Manocha.
\newblock Proxemic group behaviors using reciprocal multi-agent navigation.
\newblock In \emph{Robotics and Automation (ICRA), 2016 IEEE International
  Conference on}, pages 292--297. IEEE, 2016.

\bibitem[Gray and Wegner(2012)]{gray2012feeling}
Kurt Gray and Daniel~M Wegner.
\newblock Feeling robots and human zombies: Mind perception and the uncanny
  valley.
\newblock \emph{Cognition}, 125\penalty0 (1):\penalty0 125--130, 2012.

\bibitem[Bera and Manocha(2015)]{berareach}
Aniket Bera and Dinesh Manocha.
\newblock {REACH}: Realtime crowd tracking using a hybrid motion model.
\newblock \emph{ICRA}, 2015.

\bibitem[Bera et~al.(2016)Bera, Kim, Randhavane, Pratapa, and
  Manocha]{beraglmp}
Aniket Bera, Sujeong Kim, Tanmay Randhavane, Srihari Pratapa, and Dinesh
  Manocha.
\newblock Glmp-realtime pedestrian path prediction using global and local
  movement patterns.
\newblock \emph{ICRA}, 2016.

\bibitem[Kim et~al.(2016)Kim, Bera, Best, Chabra, and
  Manocha]{kim2016interactive}
Sujeong Kim, Aniket Bera, Andrew Best, Rohan Chabra, and Dinesh Manocha.
\newblock Interactive and adaptive data-driven crowd simulation.
\newblock In \emph{Virtual Reality (VR)}, pages 29--38. IEEE, 2016.

\end{thebibliography}

\end{document}